\documentclass[%
article,
superscriptaddress,
twocolumn,
nofootinbib,
notitlepage,
 amsmath,amssymb,
prc,
]{revtex4-1}

\usepackage{dcolumn}
\usepackage{bm}
\usepackage{hyperref}
\usepackage{xcolor}
\usepackage[czech, polish, english]{babel}

\usepackage{graphicx}
\usepackage{epstopdf, epsfig}
\usepackage[utf8]{inputenc}
\usepackage{upgreek}
\usepackage{relsize}
\usepackage[normalem]{ulem}
\usepackage{footmisc}
\usepackage{braket}
\usepackage{xcolor}

\usepackage{ulem}
\usepackage{enumitem}
\setlist{nosep}
\allowdisplaybreaks[1]
\newcommand{\dif}{\textup d}

\newcommand{\sub}[1]{\textsubscript{#1}}
\newcommand{\super}[1]{\textsuperscript{#1}}

\begin{document}


\title{Quantitative feasibility study of sequential neutron captures using intense lasers}

\author{Vojtěch Horný}
\email{vojtech.horny@eli-np.ro}
\affiliation{LULI - CNRS; École Polytechnique, CEA; Université Paris-Saclay; UPMC Université Paris 06; Sorbonne Université, F-91128, Palaiseau Cedex, France}
\affiliation{CEA, DAM, DIF, F-91297 Arpajon, France}
\affiliation{Universit\'e Paris-Saclay, CEA, LMCE, 91680 Bruy\`eres-le-Ch\^atel, France}
\affiliation{Extreme Light Infrastructure - Nuclear Physics, Horia Hulubei National Institute for Physics and Nuclear Engineering, 30 Reactorului Street, RO-077125, Bucharest-Magurele, Romania}

\author{Sophia N. Chen}
\affiliation{Extreme Light Infrastructure - Nuclear Physics, Horia Hulubei National Institute for Physics and Nuclear Engineering, 30 Reactorului Street, RO-077125, Bucharest-Magurele, Romania}

\author{Xavier Davoine}
\affiliation{CEA, DAM, DIF, F-91297 Arpajon, France}
\affiliation{Universit\'e Paris-Saclay, CEA, LMCE, 91680 Bruy\`eres-le-Ch\^atel, France}

\author{Laurent Gremillet}
\affiliation{CEA, DAM, DIF, F-91297 Arpajon, France}
\affiliation{Universit\'e Paris-Saclay, CEA, LMCE, 91680 Bruy\`eres-le-Ch\^atel, France}

\author{Julien~Fuchs}
\affiliation{LULI - CNRS; École Polytechnique, CEA; Université Paris-Saclay; UPMC Université Paris 06; Sorbonne Université, F-91128, Palaiseau Cedex, France}

\date{\today}

\begin{abstract}
Deciphering the conditions under which neutron captures occur in the Universe to synthesize heavy elements is an endeavour pursued since the 1950s, but that has proven elusive up to now due to the experimental difficulty of generating the extreme neutron fluxes required. It has been evoked that laser-driven (pulsed) neutron sources could produce neutron beams with characteristics suitable to achieve nucleosynthesis in the laboratory. In this scheme, the laser first generates an ultra-high-current, high-energy proton beam, which is subsequently converted into a dense neutron beam. 
Here we model, in a self-consistent manner, the transport of laser-accelerated protons through the neutron converter, the subsequent neutron generation and propagation, and finally the neutron capture reactions in a gold ($^{197}$Au) chosen as an illustrative example. Using the parameters of present-day available lasers, as well as of those foreseeable in the near future, we find that the final yield of the isotopes containing two more neutrons than the seed nuclei is negligible. Our investigation highlights that the areal density of the laser-driven neutron source is a critical quantity and that it would have to be increased by several orders of magnitude over the current state of the art in order to offer realistic prospects for laser-based generation of neutron-rich isotopes.
 \end{abstract}



\maketitle

\section{Introduction}
The development of high-flux neutron sources driven by ultraintense and ultrashort laser pulses has been a very active area of research over the past decade \cite{roth2013bright, pomerantz2014ultrashort, Gunther_NC_2022}. This interest is prompted by the unique properties of these sources compared to those produced by larger-scale nuclear fission reactors or accelerator-based spallation facilities \cite{Bauer_NIMA_2001, filges2009handbook}. In fact, the latter conventional sources are already outperformed by laser-based neutron pulses in terms of short duration, small size and huge instantaneous flux \cite{Martinez_MRE_2022}. Thanks to such features, promising applications in non-destructive and isotope-sensitive material analysis have recently been demonstrated \cite{Zimmer_NC_2022}. 

Another potentially attractive use of laser-driven, pulsed neutron beams that have been excogitated in the last few years \cite{chen2019extreme, hill2021exploring} would be to create neutron-rich isotopes, complementing traditional methods using nuclear reactors \cite{knapp2005production, lepareur2019rhenium}. It was anticipated \cite{chen2019extreme} that a specific benefit of such bright pulsed sources would represent an advance in laboratory astrophysics, potentially deepening our knowledge of the nucleosynthesis of elements heavier than iron.

In astrophysics, the creation of elements beyond iron is explained by slow and rapid neutron captures---the so-called $s$-process and $r$-process, respectively. In the $s$-process \cite{burbidge1957synthesis}, a seed nucleus captures a neutron, resulting in the formation of an isotope with an increased atomic mass. This is followed by a $\beta^-$-decay process, during which an electron and an electron antineutrino are emitted, transforming the nucleus into a higher-$Z$ nuclide. The $s$-process is characterized by its repeated occurrence during the star's lifetime, gradually producing heavier elements. In particular, this reaction sequence encounters critical junctures, known as branching points, which determine the pathway for isotopic evolution, with some isotopes undergoing neutron capture and others electron capture, thereby influencing the final elemental composition. Ultimately, the $s$-process terminates with the isotope $^{209}$Bi. Beyond this point, the subsequent neutron-capture product, $^{210}$Po, is subject to $\alpha$-decay, hence reducing the atomic number by two and cycling the element back to $^{206}$Pb.

The origin of 
about half of the elements heavier than iron and all of the elements heavier than bismuth is ascribed to the $r$-process,
through which sequential neutron captures drive the nuclides away from the valley of stability, in the vicinity of the neutron drip-line, until an equilibrium between neutron capture and photodissociation is reached \cite{cowan2021origin}. $\beta$-decays then occur, increasing the nucleus charge and allowing further neutron capture. This scenario is believed to be responsible for the synthesis of neutron-rich stable isotopes of heavy elements.
To be effective, though, the $r$-process requires that extreme neutron density condition ($>10^{20}\,\rm cm^{-3}$) be attained, such as those envisioned
in core-collapse supernovae \cite{Andrews_2020} or neutron star mergers \cite{kasen2017origin}.
An intermediate or $i$-process is also thought to operate at lower neutron densities, i.e., ranging from $10^{12}$ to $10^{16}\,\rm cm^{-3}$ \cite{cowan1977production, choplin2021intermediate}.

On Earth, the examination of the by-products of thermonuclear explosions \cite{diamond1960heavy, lutostansky2018production}, during which similarly extreme neutron fluxes could be achieved, demonstrated the effectiveness of the $r$-process in producing super-heavy elements. A theoretical reexamination of these results~\cite{zagrebaev2011production} concluded that a measurable amount of neutron-rich, long-lived super-heavy nuclei may be synthesized employing \textit{next-generation} pulsed reactors. In the same work~\cite{zagrebaev2011production}, the requirements to be met in order to produce long-lived, super-heavy elements were derived and a theoretical model was proposed to calculate the final isotope abundances. 

In this paper, we quantitatively evaluate the possibility of producing neutron-rich isotopes via sequential neutron captures using pulsed neutron sources accessible to both present-day and upcoming ultrahigh-intensity ($>10^{20}\,\rm W\,cm^{-2}$), short-pulse (fs-ps) lasers \cite{roth2013bright, kleinschmidt2018intense, Gunther_NC_2022, horny2022high}, as discussed in Section~\ref{sec:nsources}.
Our study relies on a slightly modified version of the model developed in Refs.~\cite{zagrebaev2011production, hill2021exploring}, as detailed in Section~\ref{sec:model} and the Appendix. Calculations of the isotope yields achievable by contemporary laser-plasma sources are presented in Section~\ref{sec:results}. The limitations of our approach are discussed in Section~\ref{sec:discussion}, and Section~\ref{sec:conclusion} summarizes the study.

\section{Contemporary laser-based neutron sources} 
\label{sec:nsources}

\begin{figure}
    \centering
    \includegraphics{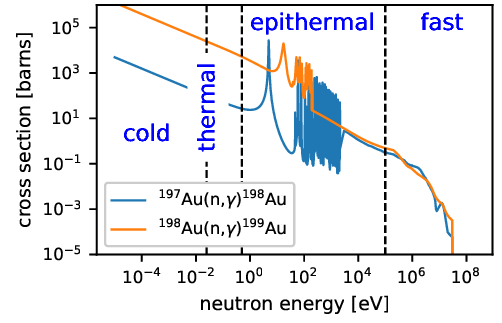}
    \caption{Neutron capture cross section dependence on energy for \super{197}Au and \super{198}Au isotopes, from the TENDL-2019 library \cite{koning2019tendl}. }
    \label{fig:FigA}
\end{figure}

Neutrons are generally categorized in terms of energy as fast ($>100$~keV), epithermal (0.5~eV -- 100~keV), thermal (25 meV), and cold ($\leq 25$~meV) \cite{mirfayzi2017experimental}. The probability of neutron capture strongly depends on the individual neutron energy; its cross-section typically grows as the neutron energy decreases as shown in Fig.~\ref{fig:FigA} for $^{197}$Au, our reference material. The plotted cross-sections are taken from the TENDL-2019 database \cite{koning2019tendl} which collects the outputs of the TALYS nuclear model code \cite{Koning_AIPC_2005}. Such an approach is necessary because measured data on neutron captures remain scarce, especially for unstable neutron-rich isotopes.

Neutron generation using intense lasers can proceed through several pathways:
(p,n) or (d,n) reactions induced by the laser-accelerated protons or deuterons on light nuclei~\cite{roth2013bright}, spallation of heavy nuclei by hundreds-of-MeV-class protons~\cite{Martinez_MRE_2022}, photonuclear reactions initiated by Bremsstrahlung photons generated by the accelerated electrons~\cite{pomerantz2014ultrashort}, or nuclear fission processes~\cite{cowan2000photonuclear}. The flux might be further enhanced, to a certain extent, by (n,2n) reaction \cite{hohenberger2022combined}. In the following, we will recall the characteristics and performances of the two main types of neutron sources based on 
laser-plasma interactions. The moderation of the energy of the generated neutrons will also be discussed.

\paragraph{Neutron sources driven by laser-accelerated fast ions}

The most common method for generating neutrons with an intense laser pulse involves high-energy electron \cite{pomerantz2014ultrashort, Gunther_NC_2022} or ion \cite{roth2013bright} beams accelerated from solid or foam targets. While the former case exploits photoreactions induced by Bremsstrahlung  $\gamma$ rays emitted by the electron beam, the latter relies on nuclear fusion reactions triggered by fast protons or deuterons on either light nuclei (e.g., Li or Be) \cite{roth2013bright} or heavy nuclei (e.g., Pb or Au) \cite{horny2022high}.

\begin{figure*}
    \centering
    \includegraphics{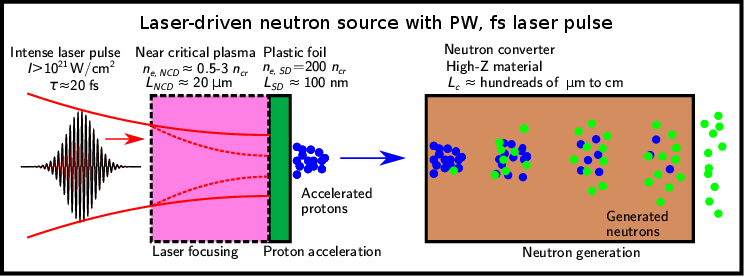}
    \caption{Example scheme for neutron generation by a laser-driven proton beam~\cite{horny2022high}. A PW-class, ultraintense ($>10^{21}\,\rm W\,cm^{-2}$), ultrashort (20~fs) laser pulse interacts with a thin (sub-micron) solid foil (green rectangle), from which protons are accelerated. To boost the on-target laser intensity, and thus proton acceleration, a foam layer (pink box), of $\sim 20\,\rm \mu m$ thickness and acting, once ionized, as a near-critical-density plasma lens, can be attached to the solid foil. The neutron converter, made of a high-$Z$ material (the brown box), is placed behind the laser target.}
    \label{fig:ldns}
\end{figure*}

Experimentally, 
100~J-class, picosecond-duration laser systems have proven effective in producing bright sources of neutrons. However, the higher intensity and repetition rate of upcoming multi-PW, few-femtosecond laser systems have the potential to enhance proton acceleration to energies over 100~MeV, where the efficiency of the spallation process in high-$Z$ nuclei drastically increases~\cite{Martinez_MRE_2022}. Such a configuration, which is considered in the following numerical study, is schematically shown in Fig.~\ref{fig:ldns}.

\begin{figure}
    \centering
    \includegraphics{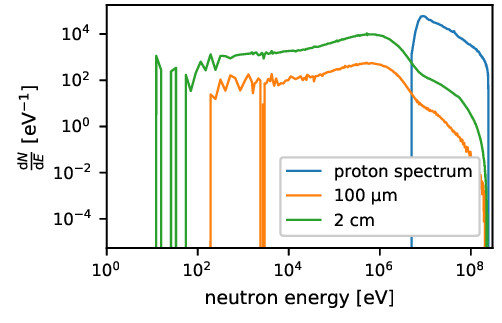}
    \caption{
    Simulated neutron energy spectra (in orange and green) for the scheme shown in Fig.~\ref{fig:ldns}, using a 1~PW, $2\times 10^{21}\,\rm cm^{-2}$, 20~fs laser pulse, according to Ref.~\cite{horny2022high}. Neutrons are generated in a Pb converter by protons (blue spectrum) accelerated from a double-layer target, composed of a 20~$\upmu$m-thick, 1~$n_c$-density carbon layer followed by a 115~nm-thick, $200\,n_c$-density CH$_2$ foil. The neutron converter has a thickness of either $100\,\rm \mu m$ (orange) or 2~cm (green). 
        } 
    \label{fig:FigB}
\end{figure}
\begin{table}
    \begin{tabular}{|l|c|c|}
\hline
	$L_c$		&	100 $\upmu$m	&	2 cm		\\
\hline
1 eV - 1 keV		&	$5.96 \times 10^4$	&	$7.64 \times 10^5$	\\
1 keV - 10 keV		&	$6.36 \times 10^5$	&	$1.16 \times 10^7$	\\
10 keV - 100 keV	&	$1.20 \times 10^7$	&	$2.10 \times 10^8$	\\
100 keV - 1 MeV		&	$4.38 \times 10^8$	&	$7.76 \times 10^9$	\\
1 MeV - 10 MeV		&	$6.64 \times 10^8$	&	$1.29 \times 10^{10}$	\\
$>$ 10 MeV	&	$8.67 \times 10^7$	&	$2.67 \times 10^9$	\\
\hline
    \end{tabular}
    \caption{Number of neutrons generated in various energy bands, as computed from the neutron spectra of Fig.~\ref{fig:FigB}, obtained from Pb converters of two different thicknesses. 
    } 
    \label{tab:tab1}
\end{table}

Figure~\ref{fig:FigB} displays typical proton (blue) and neutron (orange, green) energy spectra, as simulated using the setup of Fig.~\ref{fig:ldns}.
The laser-induced proton acceleration is described in 2D3V geometry (2D in space, 3D in momentum) with the particle-in-cell \textsc{calder} code~\cite{lefebvre2003electron}. The shown proton spectrum is that obtained from a double-layer target (a carbon plasma layer of 20-$\upmu$m thickness and 1-$n_c$ density, attached to a 115~nm, $200\,n_c$ CH$_2$ foil)~\cite{horny2022high} irradiated by a 1~PW, $2\times 10^{21}\,\rm cm^{-2}$, 20~fs laser pulse. Neutron generation in the Pb converter (of variable thickness $L_c$) is treated by the Monte Carlo \textsc{fluka} code
\cite{bohlen2014fluka, vlachoudis2009flair}. 

The orange curve corresponds to a $L_c=100$~$\upmu$m thick Pb converter: while yielding the highest peak neutron intensity (i.e., exceeding $10^{23}\,\rm n\,cm^{-2}\,s^{-1}$), this case generates only $7.6\times 10^8$ neutrons and an on-axis flux of $8.6\times 10^6\,\rm n\,sr^{-1}$ \cite{horny2022high}. For the same input proton spectrum, the neutron flux is maximized with a $L_c=2\,\rm cm$ converter: it then reaches a value of $7.5\times 10^9\,\rm n\,sr^{-1}$, for a total number of $2.4\times 10^{10}$ neutrons (see the green curve in Fig.~\ref{fig:FigB}). Considering a laser repetition rate of one shot per minute, this would translate into an averaged neutron flux of $\sim 10^8\,\rm n\,sr^{-1}$ \cite{horny2022high}.

Table~\ref{tab:tab1} presents the number of neutrons produced in various energy bands for the two considered converter thicknesses. One can see that most of the neutrons are emitted at high energies ($0.1-10\,\rm MeV$). It has also been found that the fastest neutrons ($\sim 10-100\,\rm MeV$) are directional, with an angular divergence decreasing with increasing energy, whereas neutrons below $\sim 1\,\rm MeV$ have a more or less isotropic distribution \cite{horny2022high}. The downside of such a  high mean neutron energy is a weak cross-section of neutron capture by $^{197}$Au atoms, i.e., of the order of $0.1\,\rm b$ (in$^{197}$Au) or even lower (see Figure \ref{fig:FigA}). The potential of generating neutron-rich isotopes using this kind of neutron source is discussed in Section~\ref{sec:results}.

\paragraph{Laser-driven thermonuclear sources}

Still using ultra-high power, short laser pulses, an alternative approach to generate neutrons is to exploit thermonuclear reactions in laser-heated deuterated plasmas. Numerical simulations of the interaction of 500~J, 100~fs laser pulses focused at a moderate intensity of $10^{18}$~W\,cm\textsuperscript{-2} on 1~$\upmu$m thick solid CD\textsubscript{2} foils~\cite{wu2020neutron} suggest that about $10^9$ neutrons can be generated per laser shot, having a Gaussian spectrum with $\sim 2.5\,\rm MeV$ mean energy and $\sim 0.1\,\rm MeV$ energy spread. The angular distribution of such a source is isotropic and characterized by a $\sim 10^7\,\rm n\,sr^{-1}$ flux per shot. These numbers are one or two orders of magnitude lower than those obtained using the above ion-beam-induced sources. 

On a similar scale, $\sim 3.5\times10^9$ thermonuclear neutrons have been reported from laser-driven, spherically convergent plasmas with a ns, kJ-class laser~\cite{ren2017neutron}. 
Higher-flux neutron pulses can, of course, be achieved in inertial fusion implosions at MJ-class facilities. The current published record from NIF is a total neutron yield of $\sim 2\times 10^{16}$ neutrons~\cite{zylstra2021record}, associated with a $\sim 1.6\times 10^{15}\,\rm n\,sr^{-1}$ flux per shot which can be repeated once per day. However, the scale, purpose, and repetition rate of those devices are obviously inadequate for any credible application related to neutron science. 

\paragraph{Moderation}
All the aforementioned sources deliver fast neutrons  (see Figure~\ref{fig:FigB}), yet for some applications, it may be desirable to decelerate them to lower energies using a moderator. This is done via successive elastic and inelastic collisions where energy exchanges between the neutrons and the moderating medium are dependent on the scattering angle. The materials used for this purpose are usually selected based on their moderating power, which is determined by the average logarithmic energy loss and the macroscopic scattering cross section \cite{stacey2018nuclear}. A high moderation efficiency necessitates not only a high moderating power but also a weak neutron absorption. For this reason, hydrogenous-based moderators (e.g. H\textsubscript{2}, CH\textsubscript{4}) are generally favoured. To slow the neutrons down to epithermal energies, H$_2$O, D$_2$O or high-density polyethylene moderators can be employed \cite{mirfayzi2020miniature}. Recently, moderation of laser-generated neutrons down to the cold energy range has been demonstrated via a staged scheme comprising a polyethylene pre-moderator and a cryogenic H$_2$ main moderator \cite{mirfayzi2020proof}.

\section{Modeling method}
\label{sec:model}

We now consider the irradiation of a sample of stable seed nuclei (of mass number $A$) by a short neutron pulse. The final relative abundances $b_i$ of the $(A+i)$ isotopes can be estimated using the following formulas \cite{zagrebaev2011production, chen2019extreme, hill2021exploring} (see also the Appendix):
\begin{align}
b_1 &=\frac{N_1}{N_0^0}=\mu_1\exp(-\mu_1)\simeq \mu_1 \label{eq:b1} \,,\\ 
b_2 &=\frac{N_2}{N_0^0}=\frac{\mu_1\mu_2}{2}\exp(-\mu_2)\simeq  \frac{\mu_1\mu_2}{2} \label{eq:b2} \,,
\end{align}
where
\begin{equation}
\mu_i = \frac{\dif N_n}{\dif S}\sigma_i \,.
\label{eq:mu}
\end{equation}
We have introduced $\dif N_n/\dif S$ the areal density of the neutron flux, $\sigma_i$ the cross-section of neutron capture by the $(A+i-1)$ isotope, $N_0^0$ the initial number of seed nuclei, and $N_i$ the final number of $(A+i)$ isotopes. 

We now generalize this calculation to the case of a space-dependent, energy-distributed neutron flux (as in Eqs. (1) and (2), the neutron flux is too short for the temporal dependence to play a role). Taking account of the energy dependence of $\sigma_i$, and assuming axisymmetry, the spatial map of $\mu_i(x,r)$ is then given by
\begin{equation}
    \mu_i(x,r)=\int \frac{\dif^2 N_n}{\dif S \dif E}(x,r,E)\sigma_i(E) \,\dif E \,.
    \label{eq:xi1}
\end{equation}
There follows the total number of $(A+i)$ isotopes:
\begin{equation}
    N_i = \frac{\rho}{Am_p}\iint b_i(x,r)\, 2\pi r \, \dif r \dif x \,,
    \label{eq:ni}
\end{equation}
where $\rho$ is the mass density of the seed nuclei and $m_p$ is the proton mass. Hence, the abundance of the $(A+2)$ isotopes strongly increases as the neutron areal density grows and the neutron energy decreases (see Fig.~\ref{fig:FigA}).  

\begin{figure}
    \centering
    \includegraphics{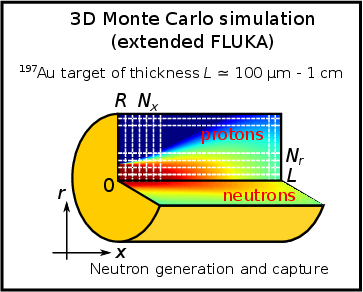}
    \caption{Setup of Monte Carlo simulations of proton transport, neutron generation and neutron transport. The protons are injected through the left-hand side of the $^{197}$Au target. The spatial grid over which neutron captures are computed is shown as white dashed lines.}
    \label{fig:scheme}
\end{figure}

To validate our numerical resolution of the above equations, we have reproduced the calculation of Ref.~\cite{hill2021exploring}, i.e. using a collimated, 5-$\rm \mu m$-diameter, a uniform cylindrical beam of $N_n=10^{12}$ neutrons, with a Gaussian energy spectrum of 0.1~MeV central energy and 10\% relative width. For a 100-$\rm \upmu m$-thick $^{176}$Lu target, we have obtained $N_1 \simeq 2.96\times 10^8$ and $N_2 \simeq 615$, consistent with the results of \cite{hill2021exploring}.

We will now consider a more realistic space-energy-distributed neutron flux based on 3D Monte Carlo \textsc{fluka} simulations, as sketched in Fig.~\ref{fig:scheme}. The protons injected into the $^{197}$Au material are either taken in the form of a uniform cylindrical beam with no initial divergence or extracted from the optimal PIC simulation case \#3 from \cite{horny2022high}. The latter describes a $2\times 10^{21}\,\rm W\,cm^{-2}$ peak intensity, 20~fs FWHM duration, $5\,\rm \upmu m$ FWHM diameter laser pulse hitting at normal incidence a double-layer target composed of a $20.5\,\rm \upmu m$-thick, near-critical ($n_e=1.05\,n_c$) carbon layer attached to a 115-nm-thick CH$_2$ foil ($n_e=200\,n_c$).
A virtual detector is placed $\sim 26\,\rm \upmu m$ behind the rear side of the target. A total of $\sim 1.5\times 10^{12}$ protons are collected (the time of arrival, transverse position and 3D momentum of each of them being recorded), characterized by an exponentially decreasing energy spectrum, extending to $\sim 240\,\rm MeV$. A module was added to the standard version of \textsc{fluka} to use as input the PIC proton distribution.

The target is a $^{197}$Au cylinder of length $L$ (along the $x$-axis) and radius $R$. The computational grid over which the neutron flux $\dif^2 N_n (x,r,E)/\dif S \dif E$ is discretized consists of $N_r$ equidistant shells (along $r$) and $N_x$ (along $x$) equidistant planes. The neutron energy, resolved by a two-way fluence (see the USRBDX entry in the \textsc{fluka} documentation \cite{bohlen2014fluka}), is tracked across all the interfaces between the grid cells and also between the outer grid cells and vacuum. 

\section{Numerical results}
\label{sec:results}

To maximize the $(A+2)$ isotope generation rate, which scales as the square of the neutron areal density, it is preferable that the stages of neutron generation and neutron captures take place within the same target. This configuration is particularly favourable when using PW, fs-class laser systems, because the resultant $> 100\,\rm MeV$ proton beams can enable efficient neutron generation in high-$Z$ materials \cite{Martinez_MRE_2022} in which neutron captures are also the most efficient. This single-target approach allows the effective neutron areal density to be significantly enhanced in the neutron-capture material. It contrasts with present-day experiments \cite{roth2013bright, kleinschmidt2018intense, mirfayzi2020proof} where neutrons are generated in cm-scale Be or Li converters, entailing a sizable drop in neutron fluence on the neutron-capture material located downstream. As an illustrative example, gold is here chosen to make up this multipurpose target: not only is it a sufficiently high-$Z$ material for neutron generation to be efficient, but is also abundant and has only one stable isotope \textsuperscript{197}Au.

In the following, we estimate the total neutron-rich isotope yield for various parameters of the accelerated protons or moderated neutrons that are injected into the neutron capture target. Seven configurations are considered as summarized in Table~\ref{tab:par}. Six of them (cases A-F) represent idealized sources of laser-accelerated protons, taken to be perfectly collimated, monoenergetic (with energies ranging from 20~MeV and 1~GeV) and with a uniform cylindrical profile of $25\,\rm \upmu m^{2}$ cross-sectional area. Based on Ref.~\cite{hill2021exploring}, the number of incident protons is set to $10^{12}$, yielding an areal density of $4 \times 10^{18}\,\rm p \,cm^{-2}$ at the target front side. The target parameters $R$ and $L$ (see Table \ref{tab:par}) are chosen such that the volume where neutron captures mainly occur is properly resolved. 

\begin{table*}[]
    \centering
\begin{tabular}{|l|ll|c|l|l|l|l|}
\hline
    & \multicolumn{1}{l|}{$E$}            & $S$             & $N$                  & $R$          & $L$    & $N_1$                & $N_2$                \\ \hline
A    & \multicolumn{1}{l|}{20 MeV}        & 25 $\upmu$m$^2$ & $10^{12}$            & 20~$\upmu$m  & 0.5~mm & $3.2 \times 10^4$    & $3.5 \times 10^{-8}$ \\ 
B    & \multicolumn{1}{l|}{50 MeV}        & 25 $\upmu$m$^2$ & $10^{12}$            & 20~$\upmu$m  & 0.5~mm & $1.5 \times 10^5$    & $4.5 \times 10^{-7}$ \\ 
C    & \multicolumn{1}{l|}{100 MeV}        & 25 $\upmu$m$^2$ & $10^{12}$            & 20~$\upmu$m  & 0.5~mm & $1.8 \times 10^5$    & $7.3 \times 10^{-7}$ \\ 
D    & \multicolumn{1}{l|}{200 MeV}        & 25 $\upmu$m$^2$ & $10^{12}$            & 15~$\upmu$m  & 0.75~mm & $2.4 \times 10^5$    & $1.0 \times 10^{-6}$ \\ 
E    & \multicolumn{1}{l|}{500 MeV}        & 25 $\upmu$m$^2$ & $10^{12}$            & 15~$\upmu$m  & 1.0~mm & $3.4 \times 10^5$    & $2.0 \times 10^{-6}$ \\ 
F    & \multicolumn{1}{l|}{1 GeV}        & 25 $\upmu$m$^2$ & $10^{12}$            & 15~$\upmu$m  & 1.0~mm & $4.0 \times 10^5$    & $2.9 \times 10^{-6}$ \\ \hline
G    & \multicolumn{2}{c|}{as case \#3 in \cite{horny2022high} }                        & $1.5 \times 10^{12}$ & 250~$\upmu$m & 0.3 mm & $4.7 \times 10^5$    & $8.5 \times 10^{-8}$ \\ \hline
\end{tabular}
    \caption{Parameters of the neutron-capture calculations: $E$, $S$ and $N$ are the energy, cross-sectional area and particle number characterizing the proton source, $R$ and $L$ are the radius and length of the neutron-capture $^{197}$Au target, and $N_1$ and $N_2$ denote the $^{198}$Au and $^{199}$Au yields. Cases A-F consider a monoenergetic, cylindrical proton beam while case G uses the proton distribution extracted from PIC simulation \#3 in \cite{horny2022high}. The values of $R$ and $L$ are chosen so as to accommodate the production of the vast majority of isotopes using a reasonable grid resolution, with the constraint that the number of internal detectors in \textsc{fluka} cannot exceed 1100 \cite{battistoni2015overview}.
    }
    \label{tab:par}
\end{table*}

\begin{figure*}
    \includegraphics{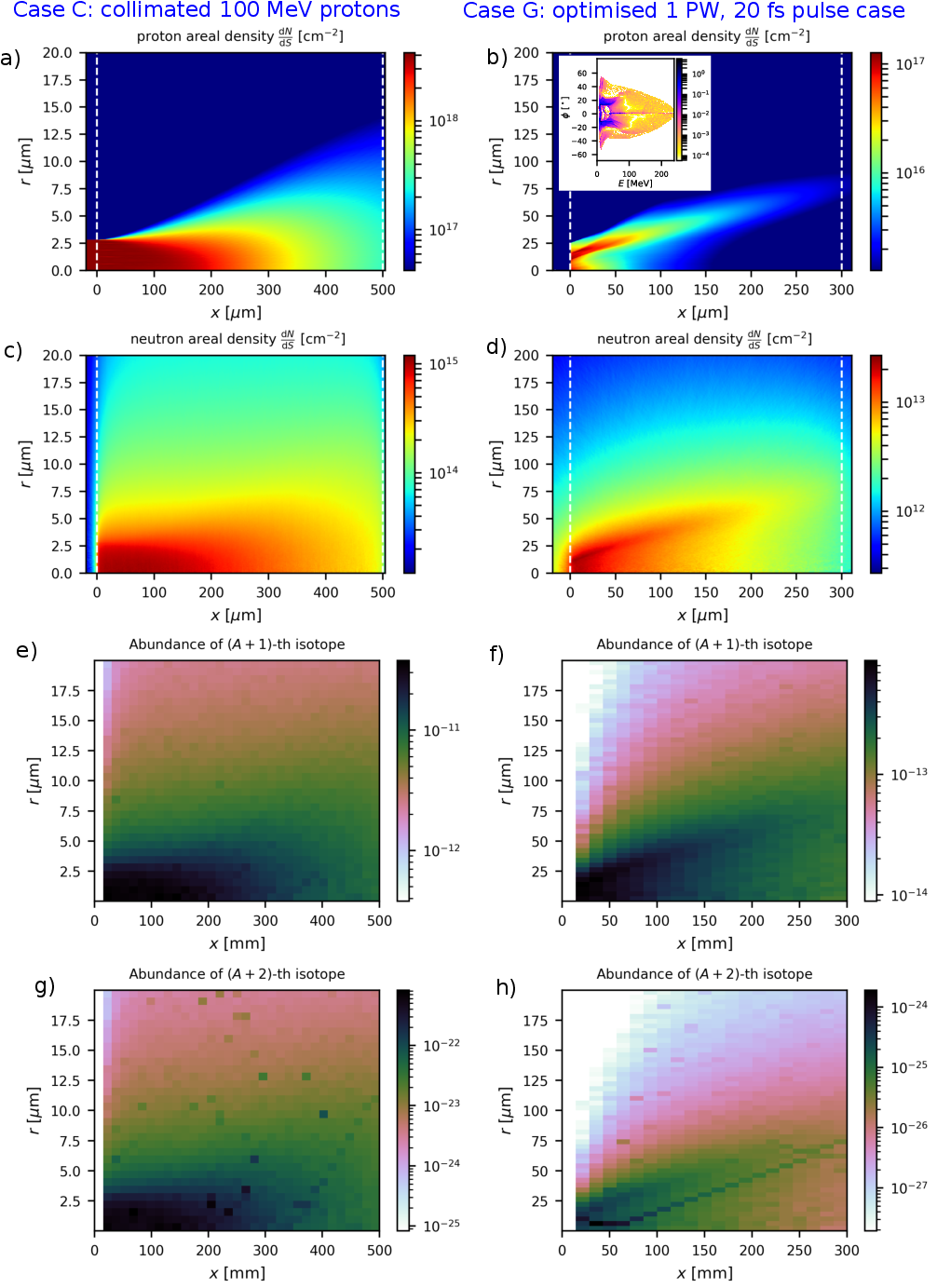}
    \caption{Proton (a-b) and neutron (c-d) transport through the $^{197}$Au converter (located between the two white dashed lines) and final abundances of \super{198}Au (e-f) and \super{199}Au (g-h) for cases C (a, c, e, g) and G from Table~\ref{tab:par}, respectively. The inset in panel (b) displays the energy-angle spectrum of the incident protons in case G.
    }
    \label{fig:FigC}
\end{figure*}

The left column of Fig.~\ref{fig:FigC} (panels a, c, e, g) details the results obtained in case C, where a narrow beam of collimated 100~MeV protons is used as input (see Table~\ref{tab:par}). Figure~\ref{fig:FigC}(a) visualizes the transport of the proton beam through the target. A steady decrease in the proton areal density is apparent: after 500~$\upmu$m of propagation, the on-axis proton areal density drops to $2.9\times 10^{17}\,\rm p\,cm^{-2}$, mainly as a result of elastic collisions, and, to a lesser degree, absorption. Panel (c) shows the corresponding neutron areal density. Its on-axis areal density reaches a maximum of $1.20\times 10^{15}\,\rm n\,cm^{-2}$ at $x_{\rm max} \simeq 54\,\rm \upmu m$, and decreases deeper into the target although the total number of neutrons reaching a position $x > x_{\rm max}$ keeps on growing with $x<L$. This behaviour arises from the increasingly divergent proton beam caused by scattering within the target. Panels (e) and (g) show the predicted abundance maps of the $^{198}$Au and $^{199}$Au isotopes, the total numbers of which are given in Table~\ref{tab:par}. While the $^{198}$Au yield per laser shot ($N_1 = 1.8\times 10^5$) could be easily measured experimentally, the $^{199}$Au yield ($N_2 = 7.3\times 10^{-7}$) is much too low to be detected, even by cumulating a large number of shots, as would be allowed, in principle, by the 3-day half-lives of $^{198}$Au and $^{199}$Au. The other idealized configurations do not provide any significant progress in this regard.

The right column of Fig.~\ref{fig:FigC} shows the results of case G, where use is made of the proton beam predicted from PIC simulation \#3 in \cite{horny2022high}, as detailed above. Panel (b) indicates that this more realistic source, which covers a transverse area of a few $100\,\rm \upmu m^2$, is characterized by a maximum areal density of $\sim 1.1\times 10^{17}\,\rm p\,cm^{-2}$ and is rather divergent with a peak at $\sim 17^\circ$ relative to the $x$-axis. Panel (d) shows that the neutrons are preferentially emitted along this direction, yet their angular distribution is not as peaked as that of the protons. A relatively high number of neutrons are produced around the $x$-axis, even though the proton density is quite low there. The explanation is that the fastest protons are injected at small angles, as seen in the energy-angle ion spectrum shown in the inset of panel (b). 
Panels (f) and (h) show that the $(A+1)$ and $(A+2)$ isotopes are created over a much larger transverse area than in case C. The total yield of $^{198}$Au isotopes ($N_1 \simeq 4.5\times 10^5$) exceeds that achieved in cases A-F. By contrast, however, only $8.5\times 10^{-8}$ $^{199}$Au isotopes are produced in total (see Table~\ref{tab:par}). This seemingly contradictory result is explained by the different scalings of the isotope abundances $b_i$ with the neutron areal density: $b_1$ scales as $b_1 \propto \dif N_n/\dif S$ whereas $b_2 \propto \left(\dif N_n/\dif S \right)^2$. 

For all considered cases, the total number of generated \textsuperscript{199}Au nuclei is at best of the order of $10^{-6}$ per shot, or even less. Hence, we can conclude that sequential neutron captures cannot be experimentally detected, and this even under the unrealistically extreme conditions of cases A-F, which indeed seem out of reach even of future 10~PW laser systems. Considering that a minimum number of about $10^{4-5}$ of isotopes per shot is necessary for their detection, contemporary short-pulse laser-driven neutron sources appear clearly unsuited to laboratory investigations of $(A+2)$ isotope generation or $r$-process nucleosynthesis, at least for the irradiation setups considered up to now. However, we will see in the following discussion that thermalization of laser-driven neutrons to cryogenic temperatures, as already demonstrated experimentally~\cite{mirfayzi2020proof}, may generate detectable amounts of $(A+2)$ isotopes, provided that the driving PW-class lasers can operate at a high enough ($\sim 100\,\rm Hz$) repetition rate. Although not currently available, such a capability can be envisioned in the relatively near future~\cite{Sistrunk_CLEAO_2017,
Hubner_Instruments_2019, Danson_HPLSE_2019}.

\section{Discussion}
\label{sec:discussion}

First, let us point out that the generation of neutron-rich isotopes can be increased by using moderators. As equation \eqref{eq:b2} suggests, the final abundance of the $(A+2)$ isotope after a certain neutron exposure scales as $\sim \sigma_1\sigma_2 \,N_n^2/S$. Hence, neutron moderation results in two opposing effects: (1) the neutron-capture cross-section significantly grows, and (2) the neutron flux considerably decreases due to geometrical reasons and absorption. The former may be greatly beneficial for the production of neutron-rich isotopes.

In the following paragraphs, we examine whether a moderated neutron source driven by a short-pulse laser could be employed for generating of $(A+2)$-th isotopes and how demanding it would be. To this goal, we consider the parameters of the cold neutron source recently reported in Ref.~\cite{mirfayzi2020proof}. In that experiment, a 300~J, 1.2~ps laser pulse was employed to generate fast protons (up to $\sim 5\,\rm MeV$) and deuterons (up to $\sim 20\,\rm MeV$) from a 5-$\upmu$m-thick C\sub{2}D\sub{4} target via target normal sheath acceleration. Those ions were subsequently converted into neutrons in a cm-scale Be target, yielding a fast neutron flux of $\sim 10^9\,\rm n \,sr^{-1}$, quite similar to case G considered above.
These fast neutrons were then slowed down through an off-axis, wing-shaped (10~mm $\times$ 6~mm $\times$ 5~mm) polyethylene pre-moderator attached to the converter and a cryogenically cooled H\sub{2} gas serving as the main moderator. The 27-nm-thick moderator was placed within a 2-mm-thick container made of copper. As a result, $\sim 10^7$ cold neutrons were released per shot, associated with a $\sim 4\times 10^5\,\rm cm^{-2}$ areal density at the surface of the H\sub{2} moderator. In the discussion section of \cite{mirfayzi2020proof}, the authors mention that their upgraded laser system could ultimately work at 100~Hz repetition rate. Considering eight hours of uninterrupted operation, the total integrated cold neutron areal density at the converter surface could be as high as $10^{12}\,\rm cm^{-2}$.

We address the same configuration, but in which the container of the cryogenic H\sub{2} gas is made of gold instead of copper. For simplicity, we further assume that the container is spherical with 27-mm diameter and 2-mm thickness \cite{mirfayzi2017experimental}. Taking the cold neutron energy distribution to be Gaussian with central energy of 20~meV, a relative energy spread of 10\% and the aforementioned available areal density of $4\times 10^5\,\rm cm^{-2}$, we can estimate that the numbers of generated \super{198}Au and \super{199}Au would be $N_1 = 3.8 \times 10^{13}$ and $N_2 = 5.3 \times 10^5$, respectively, thus making the latter detectable. We can conclude that while it is, in principle, possible to use a short-pulse laser-driven neutron source to generate $(A+2)$ isotopes, the setup involved would not bring any advantage over small, easy-to-handle gold foils at a research reactor.

Regarding the discrepancy between our estimates and the much higher isotope yield predicted in Refs.~\cite{chen2019extreme, hill2021exploring}, we first point out the greater accuracy of our modeling approach. Not only is it based on state-of-the-art Monte Carlo simulations of the proton transport, neutron generation and propagation in the target serving both as neutron converter and neutron capture target, but it also considers a realistic input source of protons as predicted by an optimized PIC simulation. The difference between our results and those of Refs.~\cite{chen2019extreme, hill2021exploring} stems chiefly from the characteristics of the neutron pulse used to induce isotope generation. The areal density ($\dif N_n/\dif S \simeq 4\times 10^{18}\,\rm n \,cm ^{-2}$) of the perfectly collimated neutron source considered in Ref.~\cite{hill2021exploring} turns out to overestimate that of laser-driven neutrons by orders of magnitude. For instance, when taking account of a realistic proton source for a 1~PW, 20~fs laser system as is done in case G above, our approach predicts $\dif N_n/\dif S \simeq 2\times 10^{13}\,\rm n \,cm ^{-2}$. Moreover, the energy spectrum of the supposedly laser-driven source in \cite{hill2021exploring} is taken to be Gaussian with a mean energy between 50~keV and 10~MeV, and a relative width of 10\%. Such parameters are at odds with previous experimental findings \cite{roth2013bright, guler2016neutron, Gunther_NC_2022, mirfayzi2017experimental} and our own results, which show much broader neutron spectra.

Similarly, we should mention that the production of $(A+2)$ isotopes was considerably overestimated in \cite{chen2019extreme} due to the erroneous assumption that the cross-section of neutron capture keeps a constant value of $1\,\rm b$ at all neutron energies. This error stemmed from the fact that the total cross-section for all reactions involving neutrons is flat; yet this does not hold when considering only multiple neutron captures, in which case the cross-section indeed drops quickly with neutron energy. This caused the cross-section to be overestimated by one to three orders of magnitude for $100\,\rm keV - 10\,\rm MeV$ neutrons, and even more for faster neutrons. This paper, however, considered a neutron flux of $10^{24}\,\rm n\,cm^{-2}\,s^{-1}$ and 1~ns duration, corresponding to an areal density of $\sim 10^{13}\,\rm n \,cm^{-2}$, similar to that characterizing our case G. The error in the cross-section led to a $(A+2)$ isotope abundance reaching $4 \times 10^{-18}$, much above that ($ \simeq 2 \times 10^{-24}$) attained in case G, see Fig.~\ref{fig:FigC}(h).

Finally, one important aspect is the production of isotopes via \textit{unwanted} channels. In our case, for example, when the gold itself acts as a proton-to-neutron converter, other reaction channels such as p-induced impurities in gold are open. As these impurities (here mainly Bi, Pb, and Tl) are likely more abundant than the freshly produced \textsuperscript{199}Au, such reactions should be considered when analyzing
the generation of the latter isotope.

\section{Conclusion}
\label{sec:conclusion}

We have carried out detailed numerical computations to assess the feasibility of realizing multiple neutron captures in laser-based experiments, under conditions relevant to astrophysical nucleosynthesis. First, considering either idealized, monoenergetic and monodirectional, primary proton beams or a realistic proton distribution, as expected to be driven by a 1 PW-class short pulse laser \cite{horny2022high}, we have performed Monte Carlo \textsc{fluka} simulations of the generation and transport of fast neutrons through a  high-$Z$ target serving both as neutron converter and neutron-capture target. $^{197}$Au was chosen as an illustrative example. We have then processed the resultant neutron areal density maps to compute the total yields of $^{198}$Au and $^{199}$Au isotopes. At variance with the encouraging conclusions reached in the recent literature \cite{hill2021exploring, chen2019extreme}, we predict that at best the negligible amount of $10^{-6}$ $^{199}$Au isotopes can be produced per laser shot, and this even by resorting to overtly optimistic proton beam parameters.

We have then tested whether thermalization of laser-driven neutrons to cryogenic temperatures, as recently demonstrated experimentally~\cite{mirfayzi2020proof}, would allow, by increasing drastically the neutron-capture cross-section, a detectable number of $(A+2)$ isotopes to be produced. We estimate that this might be indeed possible provided a jump in technology is achieved, resulting in PW-class lasers operating at least at 100~Hz repetition rate. Much effort is being currently devoted to this goal worldwide, notably through diode-pumped amplification schemes \cite{Jourdain_MRE_2014, Sistrunk_CLEAO_2017, Hubner_Instruments_2019, Danson_HPLSE_2019, Iwamoto_HEDP_2020}.

\acknowledgments 
This work was supported by funding from the European Research Council (ERC) under the European Union’s Horizon 2020 research and innovation program (Grant Agreement No. 787539, project GENESIS), and by Grant ANR-17-CE30- 0026-Pinnacle from Agence Nationale de la Recherche. Extreme Light Infrastructure Nuclear Physics (ELI-NP) Phase II, is a project co-financed by the Romanian Government and the European Union through the European Regional Development Fund and the Competitiveness Operational Programme (1/07.07.2016, COP, ID 1334). This work was supported by the contract sponsored by the Romanian Ministry of Research and Innovation: PN 23 21 01 05, and the IOSIN funds for research infrastructures of national interest. We acknowledge GENCI-TGCC for granting us access to the supercomputer IRENE under Grants No. A0100507594 and A0110512993.


\begin{thebibliography}{10}

\bibitem{roth2013bright}
M.~Roth, D.~Jung, K.~Falk, N.~Guler, O.~Deppert, M.~Devlin, A.~Favalli,
  J.~Fernandez, D.~Gautier, M.~Geissel, {\em et~al.}, ``Bright laser-driven
  neutron source based on the relativistic transparency of solids,'' {\em
  {Physical Review Letters}}, vol.~110, no.~4, p.~044802, 2013.

\bibitem{pomerantz2014ultrashort}
I.~Pomerantz, E.~Mccary, A.~R. Meadows, A.~Arefiev, A.~C. Bernstein,
  C.~Chester, J.~Cortez, M.~E. Donovan, G.~Dyer, E.~W. Gaul, {\em et~al.},
  ``Ultrashort pulsed neutron source,'' {\em {Physical Review Letters}},
  vol.~113, no.~18, p.~184801, 2014.

\bibitem{Gunther_NC_2022}
M.~G{\"u}nther, O.~Rosmej, P.~Tavana, M.~Gyrdymov, A.~Skobliakov, A.~Kantsyrev,
  S.~Z{\"a}hter, N.~Borisenko, A.~Pukhov, and N.~Andreev, ``Forward-looking
  insights in laser-generated ultra-intense $\gamma$-ray and neutron sources
  for nuclear application and science,'' {\em {Nature Communications}},
  vol.~13, no.~1, pp.~1--13, 2022.

\bibitem{Bauer_NIMA_2001}
G.~S. {Bauer}, ``{Physics and technology of spallation neutron sources},'' {\em
  {Nuclear Instruments and Methods in Physics Research Section A: Accelerators,
  Spectrometers, Detectors and Associated Equipment}}, vol.~463, pp.~505--543,
  May 2001.

\bibitem{filges2009handbook}
D.~Filges and F.~Goldenbaum, {\em Handbook of spallation research: theory,
  experiments and applications}.
\newblock Wiley, 2009.

\bibitem{Martinez_MRE_2022}
B.~Martinez, S.~Chen, S.~Bola{\~n}os, N.~Blanchot, G.~Boutoux, W.~Cayzac,
  C.~Courtois, X.~Davoine, A.~Duval, V.~Horny, {\em et~al.}, ``Numerical
  investigation of spallation neutrons generated from petawatt-scale
  laser-driven proton beams,'' {\em {Matter Radiation Extremes}}, vol.~7,
  no.~2, p.~024401, 2022.

\bibitem{Zimmer_NC_2022}
M.~{Zimmer}, S.~{Scheuren}, A.~{Kleinschmidt}, N.~{Mitura}, A.~{Tebartz},
  G.~{Schaumann}, T.~{Abel}, T.~{Ebert}, M.~{Hesse}, {\c{S}}.~{Z{\"a}hter},
  S.~C. {Vogel}, O.~{Merle}, R.-J. {Ahlers}, S.~{Duarte Pinto}, M.~{Peschke},
  T.~{Kr{\"o}ll}, V.~{Bagnoud}, C.~{R{\"o}del}, and M.~{Roth}, ``{Demonstration
  of non-destructive and isotope-sensitive material analysis using a
  short-pulsed laser-driven epi-thermal neutron source},'' {\em Nature
  Communications}, vol.~13, p.~1173, Mar. 2022.

\bibitem{chen2019extreme}
S.~Chen, F.~Negoita, K.~Spohr, E.~d’Humi{\`e}res, I.~Pomerantz, and J.~Fuchs,
  ``Extreme brightness laser-based neutron pulses as a pathway for
  investigating nucleosynthesis in the laboratory,'' {\em {Matter and Radiation
  at Extremes}}, vol.~4, no.~5, p.~054402, 2019.

\bibitem{hill2021exploring}
P.~Hill and Y.~Wu, ``Exploring laser-driven neutron sources for neutron capture
  cascades and the production of neutron-rich isotopes,'' {\em {Physical Review
  C}}, vol.~103, no.~1, p.~014602, 2021.

\bibitem{knapp2005production}
F.~Knapp~Jr, S.~Mirzadeh, A.~Beets, and M.~Du, ``{Production of therapeutic
  radioisotopes in the ORNL High Flux Isotope Reactor (HFIR) for applications
  in nuclear medicine, oncologyand interventional cardiology},'' {\em {Journal
  of Radioanalytical and Nuclear Chemistry}}, vol.~263, pp.~503--509, 2005.

\bibitem{lepareur2019rhenium}
N.~Lepareur, F.~Lac{\oe}uille, C.~Bouvry, F.~Hindr{\'e}, E.~Garcion,
  M.~Ch{\'e}rel, N.~Noiret, E.~Garin, and F.~R. Knapp~Jr, ``{Rhenium-188
  labeled radiopharmaceuticals: current clinical applications in oncology and
  promising perspectives},'' {\em {Frontiers in Medicine}}, vol.~6, p.~132,
  2019.

\bibitem{burbidge1957synthesis}
E.~M. Burbidge, G.~R. Burbidge, W.~A. Fowler, and F.~Hoyle, ``Synthesis of the
  elements in stars,'' {\em {Reviews of Modern Physics}}, vol.~29, no.~4,
  p.~547, 1957.

\bibitem{cowan2021origin}
J.~J. {Cowan}, C.~{Sneden}, J.~E. {Lawler}, A.~{Aprahamian}, M.~{Wiescher},
  K.~{Langanke}, G.~{Mart{\'\i}nez-Pinedo}, and F.-K. {Thielemann}, ``{Origin
  of the heaviest elements: The rapid neutron-capture process},'' {\em Reviews
  of Modern Physics}, vol.~93, p.~015002, Jan. 2021.

\bibitem{Andrews_2020}
S.~{Andrews}, C.~{Fryer}, W.~{Even}, S.~{Jones}, and M.~{Pignatari}, ``{The
  Nucleosynthetic Yields of Core-collapse Supernovae: Prospects for the Next
  Generation of Gamma-Ray Astronomy},'' {\em {Astrophysical Journal}},
  vol.~890, no.~1, p.~35, 2020.

\bibitem{kasen2017origin}
D.~Kasen, B.~Metzger, J.~Barnes, E.~Quataert, and E.~Ramirez-Ruiz, ``Origin of
  the heavy elements in binary neutron-star mergers from a gravitational-wave
  event,'' {\em Nature}, vol.~551, no.~7678, pp.~80--84, 2017.

\bibitem{cowan1977production}
J.~J. Cowan and W.~K. Rose, ``{P}roduction of c-14 and neutrons in red
  giants,'' {\em {Astrophysical Journal, Part 1}}, vol.~212, pp.~149--158,
  1977.

\bibitem{choplin2021intermediate}
A.~Choplin, L.~Siess, and S.~Goriely, ``{The intermediate neutron capture
  process-I. Development of the i-process in low-metallicity low-mass AGB
  stars},'' {\em Astronomy \& astrophysics}, vol.~648, p.~A119, 2021.

\bibitem{diamond1960heavy}
H.~Diamond, P.~Fields, C.~Stevens, M.~Studier, S.~Fried, M.~Inghram, D.~Hess,
  G.~Pyle, J.~Mech, W.~Manning, {\em et~al.}, ``{Heavy isotope abundances in
  Mike thermonuclear device},'' {\em {Physical Reviews}}, vol.~119, no.~6,
  p.~2000, 1960.

\bibitem{lutostansky2018production}
Y.~S. Lutostansky and V.~Lyashuk, ``Production of transuranium nuclides in
  pulsed neutron fluxes from thermonuclear explosions,'' {\em {JETP Letters}},
  vol.~107, no.~2, pp.~79--85, 2018.

\bibitem{zagrebaev2011production}
V.~Zagrebaev, A.~Karpov, I.~Mishustin, and W.~Greiner, ``Production of heavy
  and superheavy neutron-rich nuclei in neutron capture processes,'' {\em
  {Physical Review C}}, vol.~84, no.~4, p.~044617, 2011.

\bibitem{kleinschmidt2018intense}
A.~Kleinschmidt, V.~Bagnoud, O.~Deppert, A.~Favalli, S.~Frydrych, J.~Hornung,
  D.~Jahn, G.~Schaumann, A.~Tebartz, F.~Wagner, {\em et~al.}, ``{Intense,
  directed neutron beams from a laser-driven neutron source at PHELIX},'' {\em
  Physics of Plasmas}, vol.~25, no.~5, p.~053101, 2018.

\bibitem{horny2022high}
V.~Horný, S.~N. Chen, X.~Davoine, V.~Lelasseux, L.~Gremillet, and J.~Fuchs,
  ``High-flux neutron generation by laser-accelerated ions from single-and
  double-layer targets,'' {\em {Scientific Reports}}, vol.~12, no.~1, p.~19767,
  2022.

\bibitem{mirfayzi2017experimental}
S.~R. Mirfayzi, A.~Alejo, H.~Ahmed, D.~Raspino, S.~Ansell, L.~A. Wilson,
  C.~Armstrong, N.~M. Butler, R.~Clarke, A.~Higginson, {\em et~al.},
  ``Experimental demonstration of a compact epithermal neutron source based on
  a high power laser,'' {\em {Applied Physics Letters}}, vol.~111, no.~4,
  p.~044101, 2017.

\bibitem{koning2019tendl}
A.~Koning, D.~Rochman, J.-C. Sublet, N.~Dzysiuk, M.~Fleming, and S.~Van~der
  Marck, ``{TENDL: complete nuclear data library for innovative nuclear science
  and technology},'' {\em {Nuclear Data Sheets}}, vol.~155, pp.~1--55, 2019.

\bibitem{Koning_AIPC_2005}
A.~J. {Koning}, S.~{Hilaire}, and M.~C. {Duijvestijn}, ``{TALYS: Comprehensive
  Nuclear Reaction Modeling},'' in {\em International Conference on Nuclear
  Data for Science and Technology} (R.~C. {Haight}, M.~B. {Chadwick},
  T.~{Kawano}, and P.~{Talou}, eds.), vol.~769 of {\em American Institute of
  Physics Conference Series}, pp.~1154--1159, May 2005.

\bibitem{cowan2000photonuclear}
T.~Cowan, A.~Hunt, T.~Phillips, S.~Wilks, M.~Perry, C.~Brown, W.~Fountain,
  S.~Hatchett, J.~Johnson, M.~Key, {\em et~al.}, ``Photonuclear fission from
  high energy electrons from ultraintense laser-solid interactions,'' {\em
  {Physical Review Letters}}, vol.~84, no.~5, p.~903, 2000.

\bibitem{hohenberger2022combined}
M.~Hohenberger, S.~Kerr, C.~Yeamans, D.~Rusby, K.~Meaney, K.~Hahn, R.~Heredia,
  T.~Sarginson, B.~Blue, A.~Mackinnon, {\em et~al.}, ``{A combined MeV-neutron
  and x-ray source for the National Ignition Facility},'' {\em Review of
  Scientific Instruments}, vol.~93, no.~10, 2022.

\bibitem{lefebvre2003electron}
E.~Lefebvre, N.~Cochet, S.~Fritzler, V.~Malka, M.-M. Al{\'e}onard, J.-F.
  Chemin, S.~Darbon, L.~Disdier, J.~Faure, A.~Fedotoff, {\em et~al.},
  ``Electron and photon production from relativistic laser--plasma
  interactions,'' {\em {Nuclear Fusion}}, vol.~43, no.~7, p.~629, 2003.

\bibitem{bohlen2014fluka}
T.~B{\"o}hlen, F.~Cerutti, M.~Chin, A.~Fass{\`o}, A.~Ferrari, P.~G. Ortega,
  A.~Mairani, P.~R. Sala, G.~Smirnov, and V.~Vlachoudis, ``{The FLUKA code:
  developments and challenges for high energy and medical applications},'' {\em
  Nuclear Data Sheets}, vol.~120, pp.~211--214, 2014.

\bibitem{vlachoudis2009flair}
V.~Vlachoudis {\em et~al.}, ``{FLAIR: a powerful but user friendly graphical
  interface for FLUKA},'' in {\em Proc. Int. Conf. on Mathematics,
  Computational Methods \& Reactor Physics (M\&C 2009), Saratoga Springs, New
  York}, vol.~176, 2009.

\bibitem{wu2020neutron}
Y.~Wu, ``Neutron production from thermonuclear reactions in laser-generated
  plasmas,'' {\em {Physics of Plasmas}}, vol.~27, no.~2, p.~022708, 2020.

\bibitem{ren2017neutron}
G.~Ren, J.~Yan, J.~Liu, K.~Lan, Y.~Chen, W.~Huo, Z.~Fan, X.~Zhang, J.~Zheng,
  Z.~Chen, {\em et~al.}, ``Neutron generation by laser-driven spherically
  convergent plasma fusion,'' {\em {Physical Review Letters}}, vol.~118,
  no.~16, p.~165001, 2017.

\bibitem{zylstra2021record}
A.~Zylstra, A.~Kritcher, O.~Hurricane, D.~Callahan, K.~Baker, T.~Braun,
  D.~Casey, D.~Clark, K.~Clark, T.~D{\"o}ppner, {\em et~al.}, ``{Record
  energetics for an inertial fusion implosion at NIF},'' {\em {Physical Review
  Letters}}, vol.~126, no.~2, p.~025001, 2021.

\bibitem{stacey2018nuclear}
W.~M. Stacey, {\em Nuclear reactor physics}.
\newblock John Wiley \& Sons, 2018.

\bibitem{mirfayzi2020miniature}
S.~Mirfayzi, H.~Ahmed, D.~Doria, A.~Alejo, S.~Ansell, R.~Clarke,
  B.~Gonzalez-Izquierdo, P.~Hadjisolomou, R.~Heathcote, T.~Hodge, {\em et~al.},
  ``A miniature thermal neutron source using high power lasers,'' {\em {Applied
  Physics Letters}}, vol.~116, no.~17, p.~174102, 2020.

\bibitem{mirfayzi2020proof}
S.~Mirfayzi, A.~Yogo, Z.~Lan, T.~Ishimoto, A.~Iwamoto, M.~Nagata, M.~Nakai,
  Y.~Arikawa, Y.~Abe, D.~Golovin, {\em et~al.}, ``Proof-of-principle experiment
  for laser-driven cold neutron source,'' {\em {Scientific Reports}}, vol.~10,
  no.~1, pp.~1--8, 2020.

\bibitem{battistoni2015overview}
G.~Battistoni, T.~Boehlen, F.~Cerutti, P.~W. Chin, L.~S. Esposito,
  A.~Fass{\`o}, A.~Ferrari, A.~Lechner, A.~Empl, A.~Mairani, {\em et~al.},
  ``{Overview of the FLUKA code},'' {\em Annals of Nuclear Energy}, vol.~82,
  pp.~10--18, 2015.

\bibitem{Sistrunk_CLEAO_2017}
E.~Sistrunk, T.~Spinka, A.~Bayramian, S.~Betts, R.~Bopp, S.~Buck, K.~Charron,
  J.~Cupal, R.~Deri, M.~Drouin, A.~Erlandson, E.~S. Fulkerson, J.~Horner,
  J.~Horacek, J.~Jarboe, K.~Kasl, D.~Kim, E.~Koh, L.~Koubikova, R.~Lanning,
  W.~Maranville, C.~Marshall, D.~Mason, J.~Menapace, P.~Miller, P.~Mazurek,
  A.~Naylon, J.~Novak, D.~Peceli, P.~Rosso, K.~Schaffers, D.~Smith, J.~Stanley,
  R.~Steele, S.~Telford, J.~Thoma, D.~VanBlarcom, J.~Weiss, P.~Wegner, B.~Rus,
  and C.~Haefner, ``{All Diode-Pumped, High-repetition-rate Advanced Petawatt
  Laser System (HAPLS)},'' in {\em Conference on Lasers and Electro-Optics},
  p.~STh1L.2, Optica Publishing Group, 2017.

\bibitem{Hubner_Instruments_2019}
M.~H{\"u}bner, I.~Will, J.~K{\"o}rner, J.~Reiter, M.~Lenski, J.~T{\"u}mmler,
  J.~Hein, B.~Eppich, A.~Ginolas, and P.~Crump, ``Novel high-power, high
  repetition rate laser diode pump modules suitable for high-energy class laser
  facilities,'' {\em Instruments}, vol.~3, no.~3, p.~34, 2019.

\bibitem{Danson_HPLSE_2019}
C.~N. Danson, C.~Haefner, J.~Bromage, T.~Butcher, J.-C.~F. Chanteloup, E.~A.
  Chowdhury, A.~Galvanauskas, L.~A. Gizzi, J.~Hein, D.~I. Hillier, and et~al.,
  ``Petawatt and exawatt class lasers worldwide,'' {\em High Power Laser
  Science and Engineering}, vol.~7, p.~e54, 2019.

\bibitem{guler2016neutron}
N.~Guler, P.~Volegov, A.~Favalli, F.~E. Merrill, K.~Falk, D.~Jung, J.~L. Tybo,
  C.~H. Wilde, S.~Croft, C.~Danly, {\em et~al.}, ``{Neutron imaging with the
  short-pulse laser driven neutron source at the Trident laser facility},''
  {\em {Journal of Applied Physics}}, vol.~120, no.~15, p.~154901, 2016.

\bibitem{Jourdain_MRE_2014}
N.~Jourdain, U.~Chaulagain, M.~Havl{\'\i}k, D.~Kramer, D.~Kumar,
  I.~Majerov{\'a}, V.~Tikhonchuk, G.~Korn, and S.~Weber, ``{The L4n laser
  beamline of the P3-installation: Towards high-repetition rate high-energy
  density physics at ELI-Beamlines},'' {\em {Matter Radiation Extremes}},
  vol.~6, no.~1, p.~015401, 2021.

\bibitem{Iwamoto_HEDP_2020}
A.~{Iwamoto} and R.~{Kodama}, ``{Conceptual design of a subcritical research
  reactor for inertial fusion energy with the J-EPoCH facility},'' {\em {High
  Energy Density Physics}}, vol.~36, p.~100842, Aug. 2020.

\end{thebibliography}

\appendix*
\section{Calculation of the isotope abundances}

The number of isotopes generated upon irradiating a thin target layer by a neutron beam containing $N_n$ particles over an area $S$ is given by \cite{hill2021exploring}
\begin{equation}
    \mathbf{N}=e^{\mathbf{B}}\mathbf{N_0} = \left(\sum_{k=0}^{\infty} \frac{\mathbf{B}^k}{k!}\right) \mathbf{N_0} \label{eq:nmatrix} \,,
\end{equation}
where $\mathbf{N} = (N_0, N_1,N_2 \dots)^\mathrm{T}$ and $\mathbf{N_0} = (N_0^0,N_1^0, N_2^0 \dots)^\mathrm{T}$ are the initial and final populations of the $(A+i)$ isotopes ($i=0,1,2 \dots$). Furthermore, we have introduced
\begin{equation}
    \mathbf{B}= 
    \begin{bmatrix} 
    -\eta_0 & 0 & 0 & \dots & 0 \\
    \mu_0 & -\eta_1 & 0 & \dots & 0 \\
     0   & \mu_1 & -\eta_2  & \dots & 0 \\
    \vdots & \vdots & \vdots & \ddots & \vdots \\
    0 &   \dots & \dots & \mu_{l-1}     & 0 
    \end{bmatrix} \,,
\end{equation}
\begin{align}
    &\mu_i = \frac{N_n\sigma_{c,i}}{S} \,,\\
    &\eta_i = \frac{N_n\sigma_{tr,i}}{S} \,,
\end{align}
where $\sigma_{\mathrm{c},i}$ is the neutron-capture cross-section of the $(A+i)$ isotope and $\sigma_{\mathrm{tr},i}$ its transmutation cross-section, including all processes changing the neutron or proton number.

Considering only single and double neutron capture, we can rewrite Eq.~\eqref{eq:nmatrix} as 
\begin{equation}
    \mathbf{N} = \left(
    \mathbf{I} + \textbf{B} + \frac{\mathbf{B}^2}{2}
    \right)\mathbf{N_0}= 
    \begin{bmatrix} 
    1-\eta_0 + \frac{\eta_0^2}{2} & 0 & 0  \\
    \mu_0 \left(1-\frac{\eta_0}{2}-\frac{\eta_1}{2} \right) & 1-\eta_1 +\frac{\eta_1^2}{2}  & 0  \\
    \frac{\mu_0\mu_1}{2} & \mu_1\left(1-\frac{\eta_1}{2}\right) & 1 \,\\
    \end{bmatrix}
    \mathbf{N_0} \,.
\end{equation}
Given the initial conditions $\mathbf{N_0}=(N_0^0,0,0)^\mathrm{T}$, the  final populations are
\begin{eqnarray}
    N_0 &=& \left(1-\eta_0 + \frac{\eta_0^2}{2}\right)N_0^0 \approx N_0^0 \,,\\
    N_1 &=& \mu_0 \left(1-\frac{\eta_0}{2}-\frac{\eta_1}{2} \right)N_0 \approx \mu_0 N_0 \,,\\
    N_2 &=& \frac{\mu_0\mu_1}{2}N_0 \,.
\end{eqnarray}
\end{document}